\documentclass[draft]{agujournal2019}
\usepackage{url} %this package should fix any errors with URLs in refs.
\usepackage[inline]{trackchanges} %for better track changes. finalnew option will compile document with changes incorporated.
\usepackage{soul}
\usepackage{amsmath}
\usepackage{natbib}
\usepackage{lmodern}
\usepackage{gensymb}

\draftfalse

\newcommand{\aap}{{\it Astronomy \& Astrophysics}}

\newcommand{\araa}{{\it Annual Review of Astronomy and Astrophysics}}

\newcommand{\apj}{{\it The Astrophysical Journal}}
\newcommand{\apjl}{{\it The Astrophysical Journal Letters}}
\newcommand{\apjs}{{\it The Astrophysical Journal Supplement Series}}

\newcommand{\mnras}{{\it Monthly Notices of the Royal Astronomical Society}}

\newcommand{\solphys}{{\it Solar Physics}}

\chardef\us=`\_

\journalname{arXiv}

\begin{document}

\title{An Interpretable Machine Learning Approach to Understanding the Relationships between Solar Flares and Source Active Regions}

\authors{Huseyin Cavus\affil{1}, 
Jason T. L. Wang\affil{2}, 
Teja P. S. Singampalli\affil{2},
Gani Caglar Coban\affil{1},  
Hongyang Zhang\affil{2},
Abd-ur Raheem\affil{1},
Haimin Wang\affil{3}
}

\affiliation{1}{Department of Physics, Canakkale Onsekiz Mart University, 17110 Canakkale, Turkey}
\affiliation{2}{College of Computing, New Jersey Institute of Technology,  
Newark, NJ 07102, USA}
\affiliation{3}{Institute for Space Weather Sciences, New Jersey Institute of Technology, 
Newark, NJ 07102, USA}

\begin{keypoints}

\item Implementing the random forest algorithm as an interpretable machine learning approach to enable the binary classification of solar flares using the physical features of ARs collected from SolarMonitor.org
\item Today's value of Mount Wilson AR type and yesterday's value of Hale class are the most and the least governing features respectively
\item The difference in the number of spots in the AR comparing with the previous day,
i.e., NoS\_Difference,
has a distinctive effect for decision-making in both global and local interpretations
\end{keypoints}

\begin{abstract}
Solar flares are defined as outbursts on the surface of the Sun. They occur when energy accumulated in magnetic fields enclosing solar active regions (ARs) is abruptly expelled. Solar flares and associated coronal mass ejections are sources of space weather that adversely impact devices at or near Earth, including the obstruction of high-frequency radio waves utilized for communication and the deterioration of power grid operations. \\
Tracking and delivering early and precise predictions of solar flares is essential for readiness and catastrophe risk mitigation. This paper employs the random forest (RF) model to address the binary classification task, analyzing the links between solar flares and their originating ARs with observational data gathered from 2011 to 2021 by SolarMonitor.org and the XRT flare database. We seek to identify the physical features of a source AR that significantly influence its potential to trigger $\geq$C-class flares.
We found that the features of AR\_Type\_Today,  Hale\_Class\_Yesterday are the most and the least prepotent features, respectively. NoS\_Difference has a remarkable effect in decision-making in both global and local interpretations.
\end{abstract}

\section*{Plain Language Summary}
Solar flares arise from the emission of energy accumulated in the magnetic fields of solar ARs. However, the underlying mechanism for these outbursts remains unidentified. This study employs an interpretable machine learning approach to establish a solar flare prediction model, enabling the binary classification of flares ($\geq$C as the positive class and $<$C as the negative class) using the physical features of ARs obtained from SolarMonitor.org and flare data supplied from the XRT flare catalog with observational data from 2011 to 2021. 
Using the Random Forest (RF) algorithm, the performance metric values are acquired as recall of 0.81, precision of 0.82, accuracy of 0.74 and F1 score of 0.82.  
Our findings indicate that, the AR\_Type\_Today ($\beta$, $\gamma$ and $\gamma$$\delta$ types are effective in more than 90\% of positive classifications) is the most influential feature and the \\
Hale\_Class\_Yesterday ($\beta$, $\beta$$\gamma$ and $\beta$$\gamma$$\delta$ classes are dominant in over 85\% cases of positive classes) is the least influential factor for an AR generating a $\geq$C-class flare. 
The \\
NoS\_Difference feature, which takes values between $-13$ and $10$ in more than 90\% of positive test samples, significantly influences the decision-making process in both global and local interpretations.

\section{Introduction}
Active regions (ARs) on the solar disk are areas where the Sun's magnetic field is altered. They are often associated with sunspots and are the origin of violent eruptions such as coronal mass ejections (CMEs) and solar flares \citep{1995A&A...304..585H}.
The presence of sunspots indicates ARs. Ultraviolet and X-ray images of the Sun reveal these regions to be luminous because of the extraordinarily energetic events associated with ARs
\citep{1963ARA&A...1...59F}. 
These regions are surrounded by dramatic structures, such as coronal loops and solar prominences.

Different solar activities occur according to the open or closed magnetic field lines formed by the sunspot groups. Although open magnetic field lines cause the solar wind, sudden changes in magnetic field lines and reconnections of closed magnetic field lines cause solar eruptions such as CMEs and flares \citep{Priest_2014}. These eruptive events can initiate other physical activities, such as interplanetary shock waves
and geomagnetic storms, which affect the Earth's atmosphere, disrupt short-wave communications and adversely affect satellite and space operations
\citep{2022Univ....8...39M,2015ApJ...814...59V}.

Flares emerge in active regions \citep{2005ApJ...629.1141A},
especially near sunspots, where powerful magnetic fields penetrate the photosphere and connect the corona to the Sun. Flares are propelled by the immediate (minutes to tens of minutes) discharge of magnetic energy from the corona. 
\citet{2003SoPh..218..261A} reported that all the X-class flares and 55\% of the M-class flares were associated with 311 LASCO-observed CMEs between 1996 and 1999. In a more extensive mathematical investigation, \citet{2005JGRA..11012S05Y}
indicated that the association rate of CMEs significantly escalated with the magnitude of X-ray flares, 
rising from 20\% for C-class flares to 100\% for very large flares. They showed that all CMEs linked to X-class flares were observed by the LASCO coronagraphs, whereas 25-67\% of CMEs linked to C-class flares remained undetected. 
\citet{2012JAsGe...1..172Y} worked with 776 CME-flare-matched activities. They found that 67\% of the CMEs occurred after flare events. These findings indicate that although flares and CMEs
often occur together, there is no one-to-one correspondence
between them. Moreover, flare durations do not guarantee the association of CMEs \citep{1995A&A...304..585H}. 

\citet{Falconer_2012} provide a dual rationale for the increased flare productivity of active regions in their investigation of past flaring as a supplementary indicator of free magnetic energy to predict solar eruptions. Firstly, these ARs typically exhibit a complex multipolar configuration of opposite-polarity magnetic flux instead of a singular bipole arrangement; secondly, they are undergoing rapid evolution through convective flux transfer, flux emergence, and/or flux cancellation. Flares emit energy in a variety of ways, including electromagnetic radiation, particles (protons and electrons), and mass streams \citep{2012annG...55...49B}. They are distinguished by their X-ray brilliance (X-ray flux). X-class flares are the largest. M-class flares have one-tenth the energy of
X-class flares. C-class flares have one-tenth the energy of
M-class flares \citep{1989ARA&A..27..421B}.
\citet{2012ApJ...760...31K} and \citet{2016ApJ...820L..11J}
worked on magnetic classes of ARs for the period between 1992 and 2015. These authors found that $\alpha$ and $\beta$ containing ARs make up 20\% and 80\%, respectively. 

In their study on the development of the determined magnetic complexity of ARs for the 23rd solar cycle, \citet{10.1111/j.1365-2966.2010.16465.x} stated that the most strong flares and rapid coronal mass ejections (CMEs) typically originate in active regions (ARs) having intricate structures. \citet{MOHAMED2018249} 
performed a statistical analysis of the frequency of X-class flares during the declining and peak phases of Solar Cycles 23 and 24, respectively. They found a consistency between 
the number of X-class flares and the number of days of $\beta$$\gamma$$\delta$ group ARs.

In this study, we adopt an
interpretable machine learning
(ML) approach to examine the relationships between solar flares
and their source ARs using observational data collected from
2011 to 2021. By adopting the interpretable ML approach, we attempt to explain how the ML model makes decisions in determining the relationships
between flares and their source ARs.
This helps us to understand which physical properties of a source AR might play an important role in producing a flare.

The remainder of this paper is organized as follows.
Section \ref{sec:data} describes the data used in this study.
Section \ref{sec:methods} formalizes the problem studied here into a binary classification task and details our approach 
to solving the task. Section \ref{sec:results} reports the experimental results. Section \ref{sec:discussion} presents a discussion of the results. Section \ref{sec:conclusions} concludes the paper.

\section{Data}
\label{sec:data} 

The database at SolarMonitor.org was utilized to analyze
ARs and their magnetic group properties \citep{2002SoPh..209..171G}. 
It aggregates solar activity data from various sources, including the Solar Dynamics Observatory (SDO) with HMI and AIA instruments, the Global H$\alpha$ Network, 
the Solar and Heliospheric Observatory
(SOHO) with EIT and MDI instruments, 
Synoptic Optical Long-term Investigations of the Sun (SOLIS)
with full-disk chromospheric magnetograms, 
SECCHI, and databases from the National Oceanic and Atmospheric Administration (NOAA). 
It also features images from the Hinode XRT team, 
STEREO's Extreme UltraViolet Imager (EUVI),
and the Solar X-ray Imager (SXI),
as well as X-ray images from the X-ray SXT full-disk database. 

The SolarMonitor system provides AR data, 
flare forecasting information, 
and comprehensive solar disk images of the Sun. 
Its database offers near-real-time data on solar activity.
Specifically, the database includes the following physical properties
for an AR with NOAA number, the latest position, and
associated flares if available:
Hale class, AR type, sunspot area (millionths of the solar disk area), and number of spots.

Hale class has 8 values:
$\alpha$,
$\beta$,
$\alpha$$\gamma$,
$\alpha$$\delta$,
$\alpha$$\gamma$$\delta$,
$\beta$$\gamma$,
$\beta$$\delta$,
$\beta$$\gamma$$\delta$.
The Hale classes are then regrouped into five groups to obtain the Mount Wilson classification (AR type) with five values:
$\alpha$,
$\beta$,
$\gamma$,
$\delta$,
$\gamma$$\delta$
where
\begin{itemize}
\item 
$\alpha$ refers to Hale class $\alpha$;
\item 
$\beta$ refers to Hale class $\beta$; 
\item 
$\gamma$ refers to Hale classes $\alpha$$\gamma$ and $\beta$$\gamma$;
\item 
$\delta$ refers to Hale classes $\alpha$$\delta$ and $\beta$$\delta$;
\item 
$\gamma$$\delta$ refers to Hale classes $\alpha$$\gamma$$\delta$ and $\beta$$\gamma$$\delta$.
\end{itemize}

In summary, our study used 10 physical features, 
listed in Table \ref{tab:10features}, obtained from the SolarMonitor database.
Each AR record contains values of the 10 features
of the corresponding AR.

\begin{table}[h]
    \centering
    \begin{tabular}{|c|c|}
        \hline
        \textbf{Feature} & \textbf{Description} \\ \hline
        Hale\_Class\_Today & Hale class of the AR in today \\ \hline
        Hale\_Class\_Yesterday & Hale class of the AR 
        in yesterday\\ \hline
        AR\_Type\_Today & Mount Wilson classification of 
        the AR in today\\ \hline
        AR\_Type\_Yesterday & Mount Wilson classification of the AR in yesterday \\ \hline
        Spot\_Area\_Today & Area of spots of the AR in today \\ \hline
        Spot\_Area\_Yesterday & Area of spots of the AR in yesterday  \\ \hline
        Spot\_Area\_Difference  &  Spot\_Area\_Yesterday $-$
        Spot\_Area\_Today\\ \hline
        NoS\_Today & Number of spots of the AR in today \\ \hline
        NoS\_Yesterday & Number of spots of the AR in yesterday \\ \hline
        NoS\_Difference &  NoS\_Yesterday
        $-$ NoS\_Today \\ \hline 
    \end{tabular}
    \vspace*{+0.5cm}
    \caption{Physical Properties or Features of an AR Considered in Our Study}
    \label{tab:10features}
\end{table}

The flare data was sourced from the XRT flare catalog. 
This catalog contains extensive flare information, 
including positions, classes, GOES fluxes of the flare source regions, and the source AR's NOAA numbers. 
In this catalog, the magnetic group assignments of the flare source ARs were determined using the SolarMonitor database. The data from SolarMonitor.org and XRT were acquired using a web scraping script written in Python. 
We collected 986 AR-produced flare events in the period
between January 2011 and October 2021. 
After removing ARs with incomplete or missing feature values, 
we obtained 837 AR records, where each AR record
contains values of the 10 physical features of the corresponding AR and
the flare event produced by the AR.
Table \ref{tab:flareevents} summarizes
the types and counts of these flare events.

\begin{table}[h]
    \centering
    \begin{tabular}{|c|c|}
        \hline
        \textbf{Flare Type} & \textbf{Counts} \\ \hline
        A-Class & 1 \\ \hline
        B-Class & 231 \\ \hline
        C-Class & 539\\ \hline
        M-Class & 61\\ \hline
        X-Class & 5\\ \hline   
    \end{tabular}
        \vspace*{+0.5cm}
    \caption{Types and Counts of AR-Produced Flares Considered in Our Study}
    \label{tab:flareevents}
\end{table}

\section{Methodology}
\label{sec:methods}

\subsection{Classification Task}
\label{sec:task}

We consider $\geq$C-class flares,
which have been widely studied in
the literature
\citep{2023NatSR..1313665A,2024FrASS..1098609P,2018ApJ...856....7H,2020ApJ...891...10L,2019ApJ...877..121L,
2018ApJ...858..113N,2022ApJ...941....1S,2021ApJS..257...50T,2023SoPh..298..137V,2023MNRAS.521.5384Z}. 
A $\geq$C-class flare refers to a C-class,
M-class, or X-class flare and
a $<$C-class flare refers to an A-class or B-class flare.

Given an AR record with 10 physical properties or features
listed in Table \ref{tab:10features}, 
we want to solve the following binary classification problem:
does the corresponding AR produce a $\geq$C-class flare? 
Through binary classification, we attempt to understand
which physical properties or features of a source AR play an important role in determining whether
the AR produces a $\geq$C-class flare. 

\subsection{Interpretable Classification Model}
\label{sec:shap}

We adopted the random forest (RF) model to solve the binary classification task
described in Section \ref{sec:task}.
RF based models have been shown to be very effective in predicting solar flares
\citep{2017ApJ...843..104L}.
The hyperparameters were tuned using
the randomized search capability available in the Python machine learning library, {\tt scikit-learn}
\citep{10.5555/1953048.2078195}.
There were 100 trees in the random forest. 
For model interpretation, we adopted
the SHAP (SHapley Additive exPlanations) framework
\citep{DBLP:conf/nips/LundbergL17}.
SHAP is used to calculate the contribution of each feature to the final output. 
Let $F$ be the set of the 10 features in Table \ref{tab:10features}
and let $S$ be a subset of $F$.
The SHAP value of each feature $i$ in $F$, $1 \leq i \leq |F|$,
denoted $\phi_{i}$,
is defined as:
\begin{equation}
\phi_{i} = \sum_{S \subseteq F-\{i\}}\frac{|S|!(|F|-|S|-1)!}{|F|!}(C(S\cup\{j\})-C(S)).
\end{equation}
The SHAP value determines the difference in the contribution that the feature $i$ brings to the prediction if included in
a specific subset $S$, and averages the differences over every possible
combination of possible subsets $S$ of the features in terms of the
contribution function:
$C(S\cup\{j\})-C(S)$.
SHAP can be used for global interpretation (that is, the interpretation is made over the entire test set)
or local interpretation (the interpretation is made for a specific test sample).
We adopted {\tt shap.TreeExplainer} in
our study.

\section{Results}
\label{sec:results}

\subsection{Evaluation Metrics}

Given an AR record $R$ with 10 physical feature values,
we define $R$ as a true positive (TP) 
if our RF model predicts that $R$ belongs to the positive class
(i.e., the corresponding AR produces a $\geq$C-class flare)
and $R$ is indeed positive.
We define $R$ as a false positive (FP)
if our RF model predicts that $R$ is positive
while $R$ actually belongs to the negative class
(i.e., the corresponding AR produces a $<$C-class flare). 
We say $R$ is a true negative (TN) if our RF model
predicts that $R$ is negative
and $R$ is indeed negative; 
$R$ is a false negative (FN) if
our RF model predicts that $R$ is
negative while $R$ is actually positive.
When the context is clear, we also use
TP (FP, TN, and FN, respectively) 
to represent the number of
true positives
(false positives, true negatives, and false negatives, respectively) 
produced by our RF model.

The evaluation metrics used in this study include the
following:
\begin{equation}
    \text{Recall} = \frac{\mbox{TP}}{\mbox{TP} + \mbox{FN}},
\end{equation}

\begin{equation}
    \text{Precision} = \frac{\mbox{TP}}{\mbox{TP} + \mbox{FP}},
\end{equation}

\begin{equation}
    \text{Accuracy} = \frac{\mbox{TP} + \mbox{TN}}
    {\mbox{TP} + \mbox{FP} + \mbox{TN} + \mbox{FN}},
\end{equation}

\begin{equation}
\text{F1} =  \frac{2 \times \text{TP}}{2 \times \text{TP} + \text{FP}+\text{FN}}.
\end{equation}

We split the set of 837 AR records at hand into a
80\% training set and a 20\% test set.
There are a total of 605 AR records
in the positive class and
232 AR records in the negative class.
After splitting, the training set has 484 positive AR records and 185 negative AR records, totaling 669 training AR records or training samples.
The test set has 121 positive AR records
and 47 negative AR records, totaling 168 test AR records or test samples.
Each AR record in the training set contains 10 AR feature values together with a label
(positive vs. negative).
Each AR record in the test set contains 10 AR feature values without a label.
The label in a test sample will be predicted by our classification model.
We compute the TP, FP, TN, FN and all evaluation metric values based on the test set.

\subsection{Performance Assessment}

\begin{figure}
\centering
\hspace*{-0.7cm}
\includegraphics[width=0.4\linewidth]{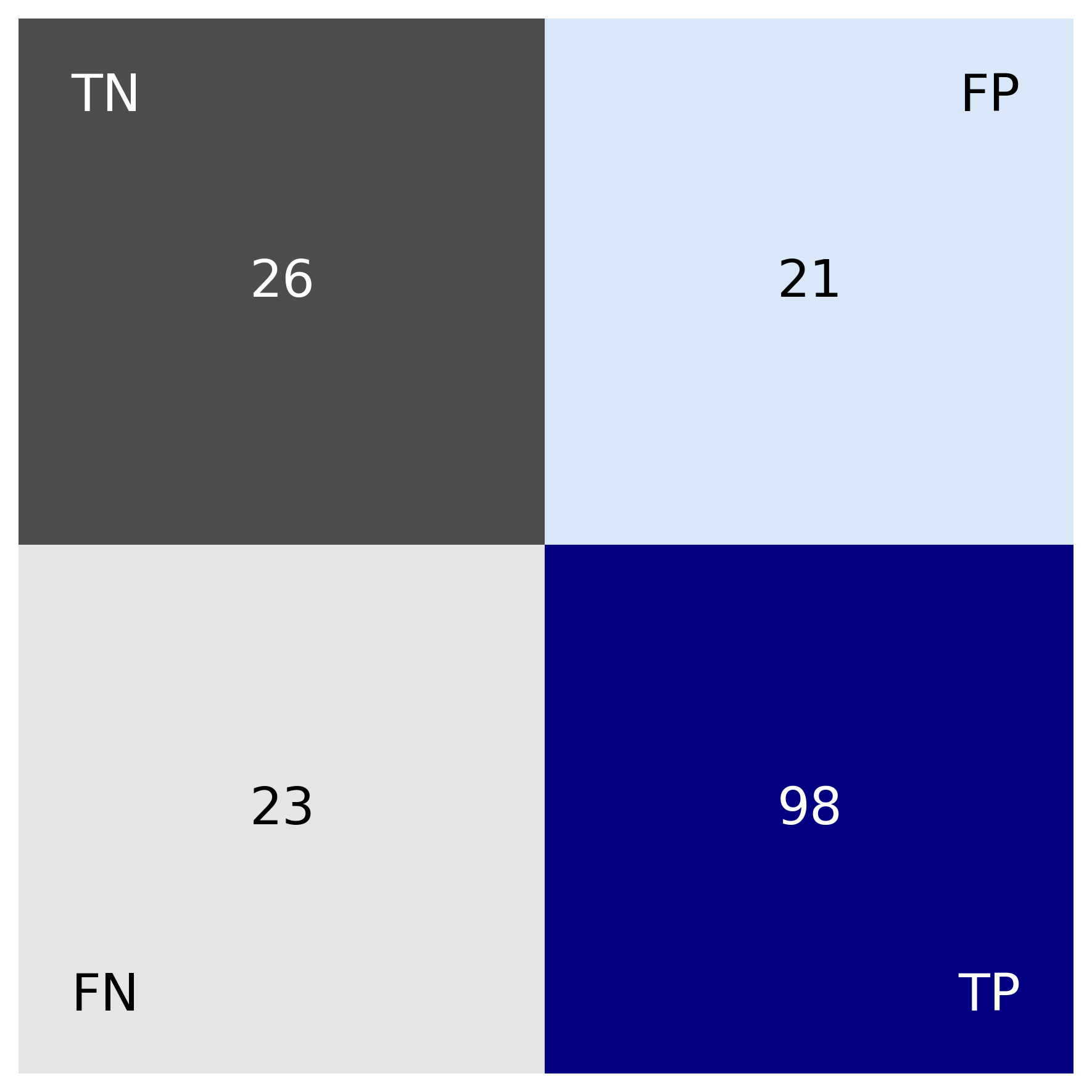}
\caption{Confusion matrix
obtained by our RF model on
the test set with 168 test samples.
}
\label{fig:confusion_matrix}
\end{figure}

Figure \ref{fig:confusion_matrix} presents the confusion matrix obtained by our RF model, which provides a breakdown
analysis of errors that occur
when the model makes predictions in the test set. 
Based on the confusion matrix, we obtain:
recall = 0.81,
precision = 0.82,
accuracy = 0.74,
F1 = 0.82.
To assess the stability and reliability of our RF model, we also performed a five-fold cross-validation.
The average accuracy for five folds is 0.75.
These results indicate that our RF model performs reasonably well in solving the binary classification problem described in
Section \ref{sec:task}.

\subsection{Model Interpretation}
\label{sec:SHAP}

As described in Section \ref{sec:shap}, 
SHAP incorporates game theory to give each feature a
SHAP value. Positive SHAP values have a positive effect,
while negative SHAP values have a negative effect. A positive
effect increases the probability of predicting that a test
sample is in the positive class, whereas a negative effect
increases the probability of predicting that a test sample is
in the negative class. 

\begin{figure}
\centering
\hspace*{-0.7cm}
\includegraphics[width=0.75\linewidth]{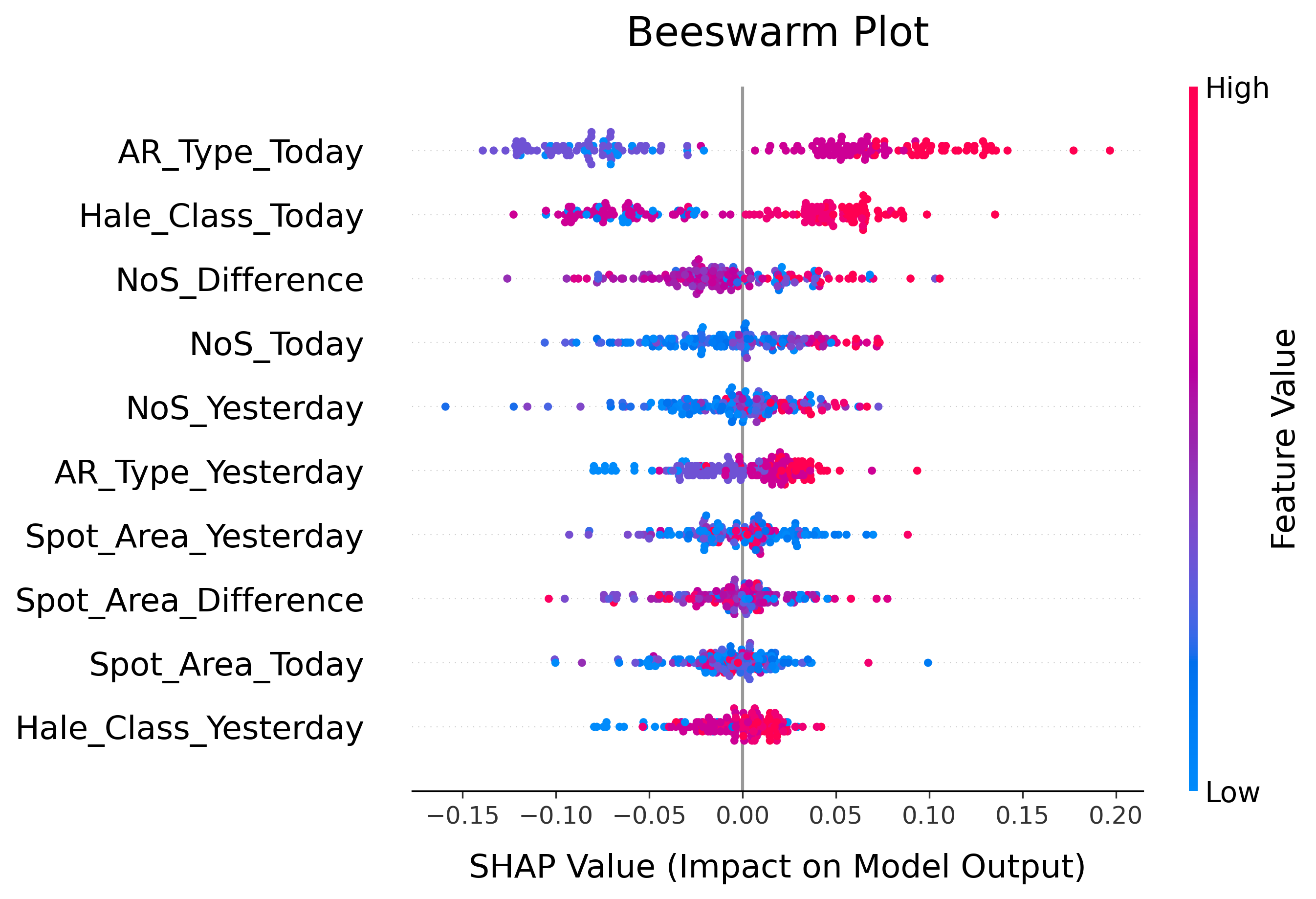}
\caption{Beeswarm plot to visualize the positive or negative
effect of a feature for each test sample, represented by a
color dot, on the RF model’s predictions.}
\label{fig:beeswarm1}
\end{figure}

Figure \ref{fig:beeswarm1} presents the beeswarm plot for the RF model. In the beeswarm plot, we can observe, for each feature, the
distribution of SHAP values for all test samples. Each test
sample corresponds to a dot in each feature row. The placement
of each dot on the $x$-axis is determined by the SHAP
value that the corresponding feature of the corresponding test sample receives. When dots cluster, it shows common test samples based on their SHAP values of the corresponding feature. 
The color of a dot depends on the value of a feature
in the corresponding test sample. Red indicates a high
feature value, blue indicates a low feature value, and purple
indicates an average or moderate feature value.

Now, focus on the NoS\_Today feature in Figure \ref{fig:beeswarm1}. There are
more test samples with negative SHAP values than with positive
SHAP values. This means that the NoS\_Today feature is
more likely to push the model’s predictions towards a negative
class. Furthermore, low feature values (blue dots) tend
to lead to negative predictions, as these low feature values
have negative SHAP values. High feature values (red dots)
tend to lead to positive predictions, as these high feature values
have positive SHAP values.

Next, focus on the Hale\_Class\_Yesterday feature in 
Figure \ref{fig:beeswarm1}.
If we compare this feature with the 
NoS\_Today feature,
this feature may appear to have fewer dots. The reason
for this appearance is that most of the SHAP values for
Hale\_Class\_Yesterday are zero or near zero. 
This causes the dots to
cluster up and makes Hale\_Class\_Yesterday seem to have fewer dots.
Furthermore, since most of the SHAP values for the feature
are zero or near zero, the Hale\_Class\_Yesterday feature neither pushes
the model’s predictions towards a negative class nor pushes
the model’s predictions towards a positive class, implying
that the Hale\_Class\_Yesterday feature plays an unimportant role in the
model’s predictions.

\begin{figure}
\centering
\hspace*{-0.7cm}
\includegraphics[width=0.75\linewidth]{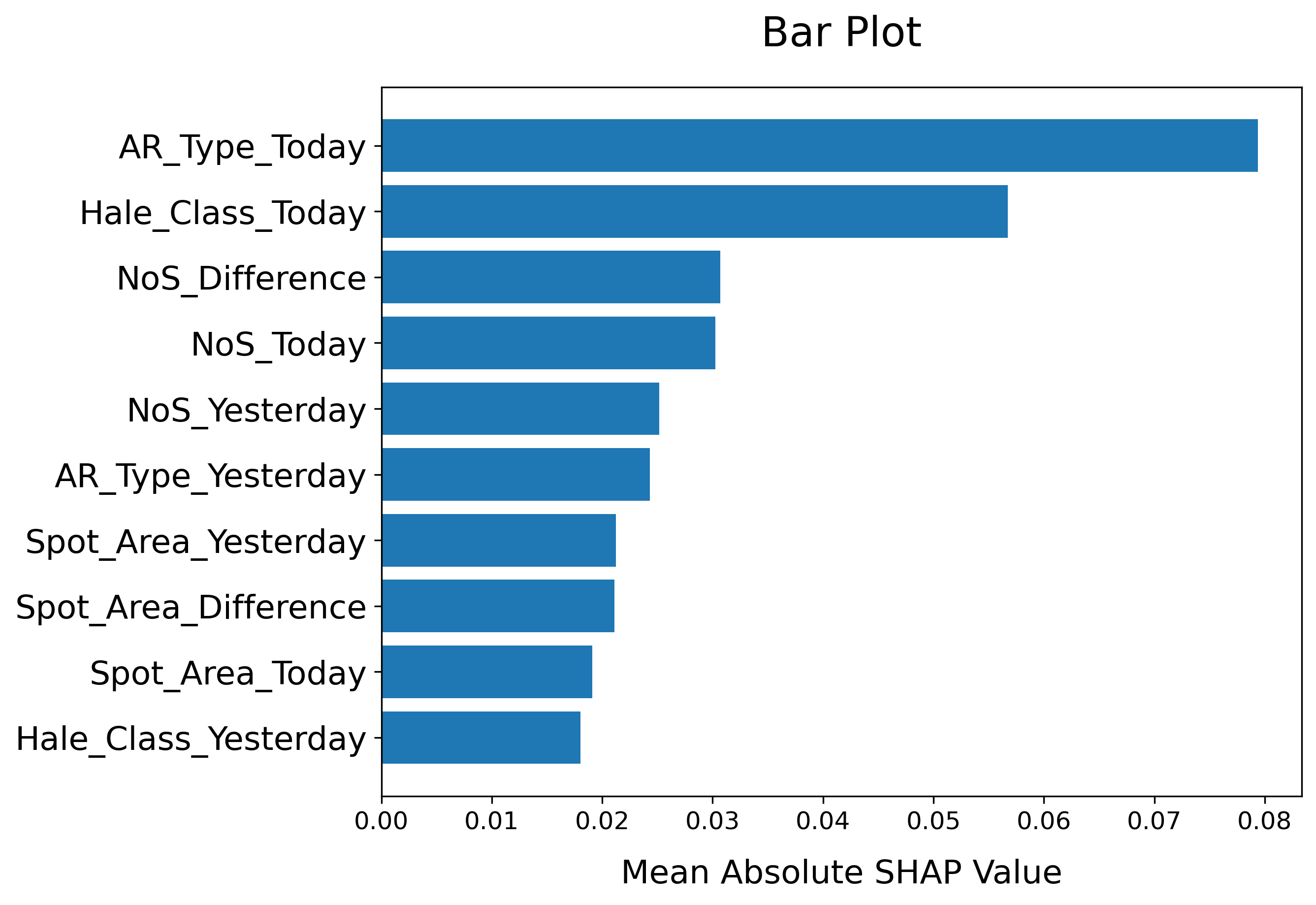}
\caption{Bar plot to display the global importance of each
feature on our RF model’s predictions.}
\label{fig:barplot}
\end{figure}

Figure \ref{fig:barplot} presents the bar plot
for the RF model. In
the bar plot, each feature is given the mean of the absolute
SHAP values across all test samples. This mean of the absolute
SHAP values is how the importance of each feature
is measured in the bar plot. The longer the bar in the bar
plot, the more important the corresponding feature is to the
model’s predictions. We can see in Figure \ref{fig:barplot} that the AR\_Type\_Today
feature is of the highest importance to the model’s predictions,
while the Hale\_Class\_Yesterday feature is of the lowest importance to the model’s predictions

In contrast to the bar plot in Figure \ref{fig:barplot}, the beeswarm plot
in Figure \ref{fig:beeswarm1} displays a separate SHAP value for each test
sample, and shows the variability of the test samples, represented
by color dots, for each feature. 
The fewer clusters
and the more spread out from the zero in the beeswarm plot
indicate higher SHAP values (positive or negative), causing
a higher mean of absolute SHAP values and ultimately a
higher importance. Now, consider again the appearance of
Hale\_Class\_Yesterday in the beeswarm plot in Figure \ref{fig:beeswarm1}. The result of
having most of its SHAP values equal or near zero reflects
how short its bar is in the bar plot in 
Figure \ref{fig:barplot}.
On the other hand, the nature of high variability of the 
AR\_Type\_Today feature
in the beeswarm plot in 
Figure \ref{fig:beeswarm1}
causes its bar to be significantly
longer than those of the other features in the bar plot
in Figure \ref{fig:barplot}.

\begin{figure}
\centering
\hspace*{-0.7cm}
\includegraphics[width=0.7\linewidth]{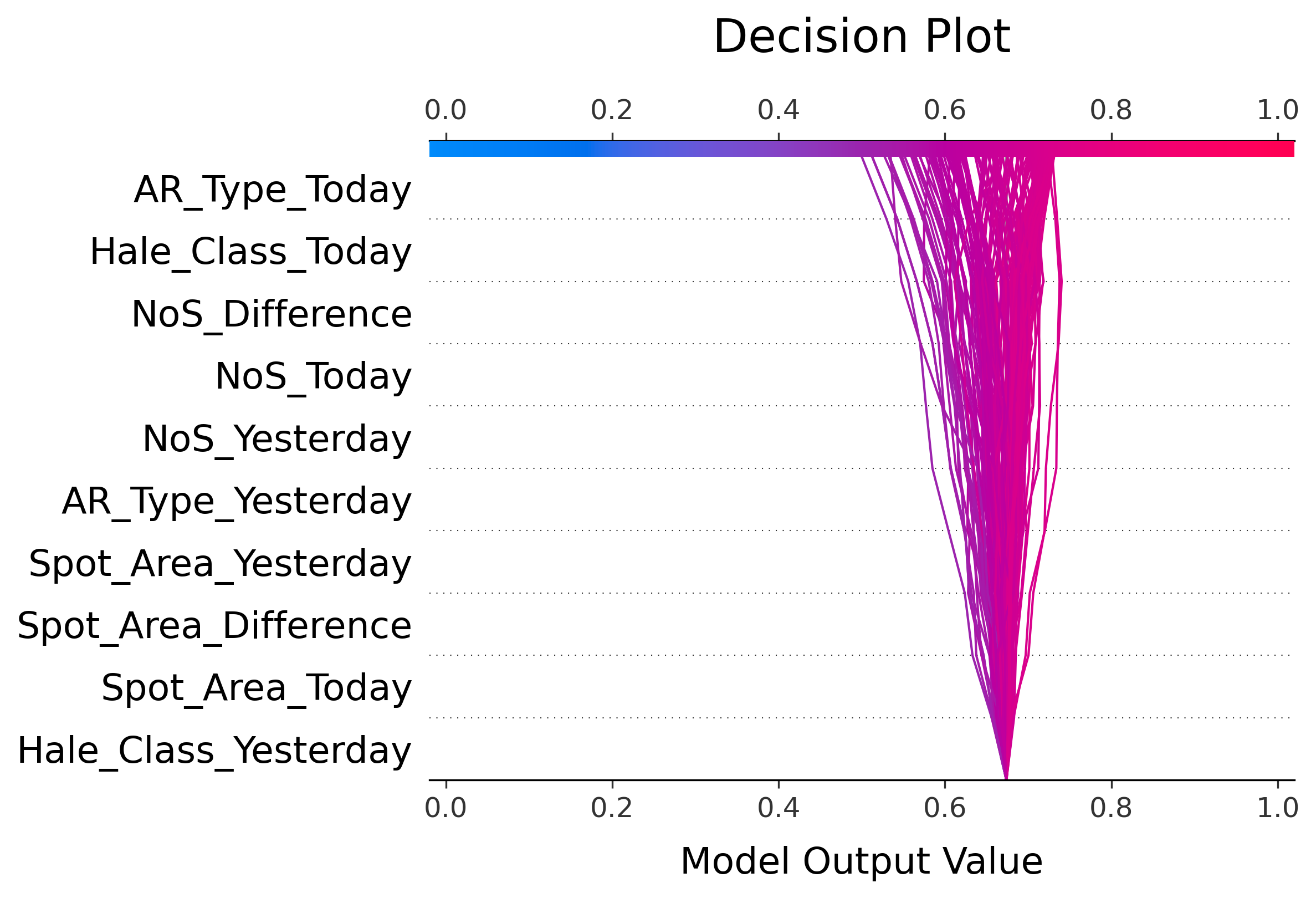}
\caption{Decision plot to understand how our RF model produces
its predictions.}
\label{fig:dec1}
\end{figure}

Figure \ref{fig:dec1} 
presents the decision plot for our RF model.
On the $y$-axis of the decision plot, the features are displayed,
from top to bottom, according to their importance, with the most
important feature displayed at the top and the least important
feature displayed at the bottom. 
Each prediction/test sample
is represented by a line in the decision plot. The predictions
for the test samples begin at the bottom with the same base
value, which is approximately 0.67.
This base value is the
mean of the model's prediction values over all training samples
in the training set. It is used as a starting point before
considering any feature contribution. 
As a prediction/line moves from bottom to top, the SHAP value for each feature
is added to the base value. This can help to understand the
contribution of each feature in the prediction. The features
can have positive or negative contributions, pushing the corresponding
line to the right or left, respectively.
The $x$-axis shows the model output values of the test samples in the test set.
The prediction value (i.e., the model output value) of a test
sample is the probability that the test sample belongs to the
positive class. 
For example, a prediction value of 0.6 indicates
that there is a 60\% chance that the test sample belongs
to the positive class.
At the top, each prediction/line reaches
its final prediction value (predicted probability). 
The prediction value determines the color of the corresponding line. A blue/purple line indicates
a lower prediction value (lower predicted probability closer to 0), 
while a red line indicates a higher prediction
value (higher predicted probability closer to 1). 

The results shown in the beeswarm, bar, and decision
plots are consistent. Based on these results, we conclude
that AR\_Type\_Today and Hale\_Class\_Today play the most 
important roles in the model's decision-making process.
The difference between the number of sunspots of an AR in today and
the number of sunspots of the AR in yesterday, i.e.,
NoS\_Difference, also plays a very important role in the 
model’s predictions.
The three plots (beeswarm, bar, and decision) are used mainly for global interpretation based on the entire test set.
In what follows, we will present waterfall plots for local interpretation,
which look at individual predictions made by
our RF model. With waterfall plots, we can better understand
each feature and why certain behaviors lead to a specific
prediction.

\begin{figure}
\centering
\hspace*{-0.7cm}
\includegraphics[width=0.8\linewidth]{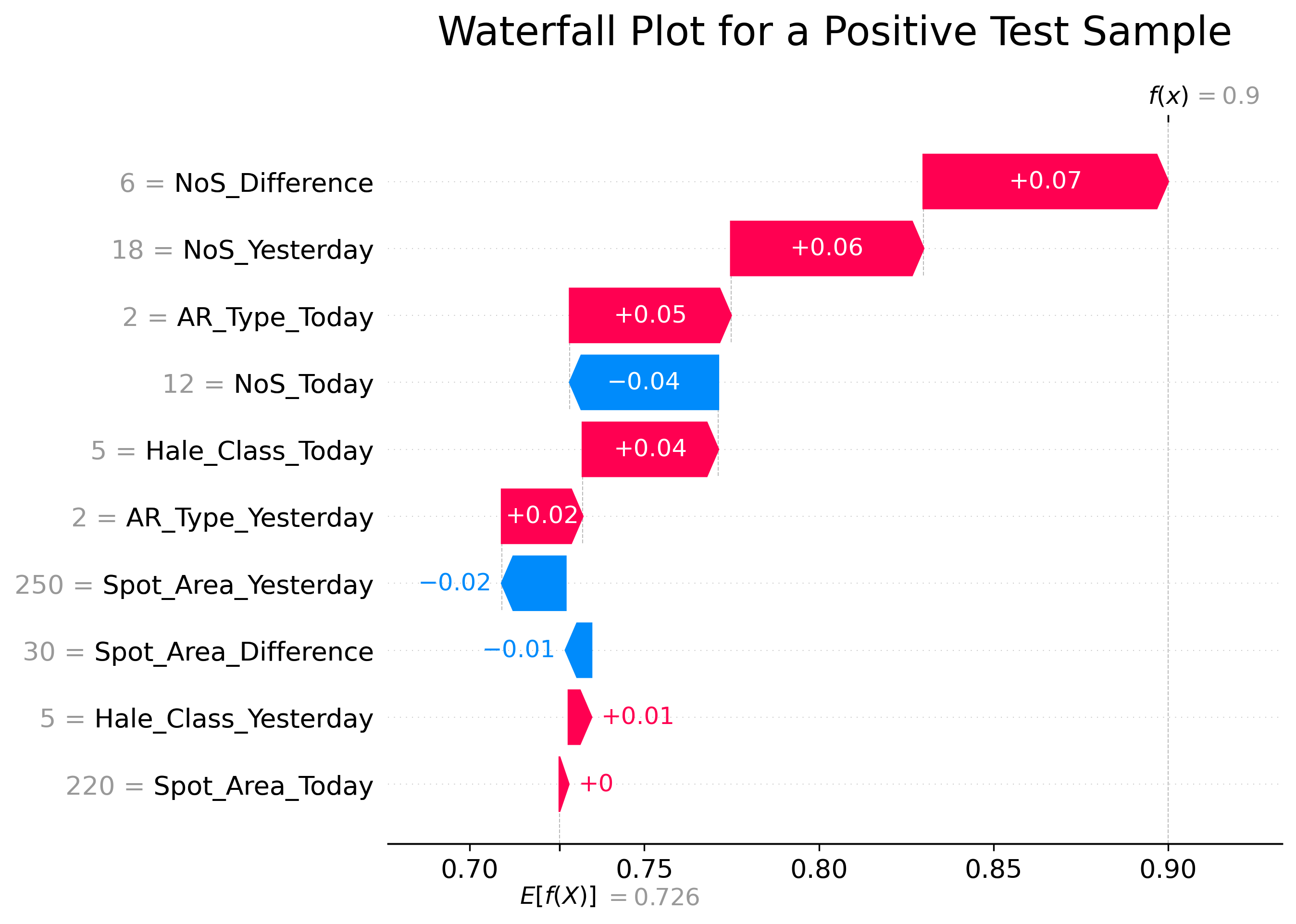}
\caption{Waterfall plot for a test sample predicted to be positive. The plot shows the relative contribution of each feature to
the model’s prediction $f(x)$ = 0.9, 
starting from the base value $E[f(x)]$ = 0.726. 
The $x$-axis represents the model output value (predicted probability)
while the $y$-axis 
shows the features and their value.
We encode the categorical features AR\_Type and Hale\_Class where AR\_Type = 2 represents $\gamma$
and
Hale\_Class = 5 represents $\beta$$\gamma$.
The arrows display the SHAP value associated with each
feature, colored red if positive and blue if negative.}
\label{fig:waterfallpositive}
\end{figure}

\begin{figure}
\centering
\hspace*{-0.7cm}
\includegraphics[width=0.8\linewidth]{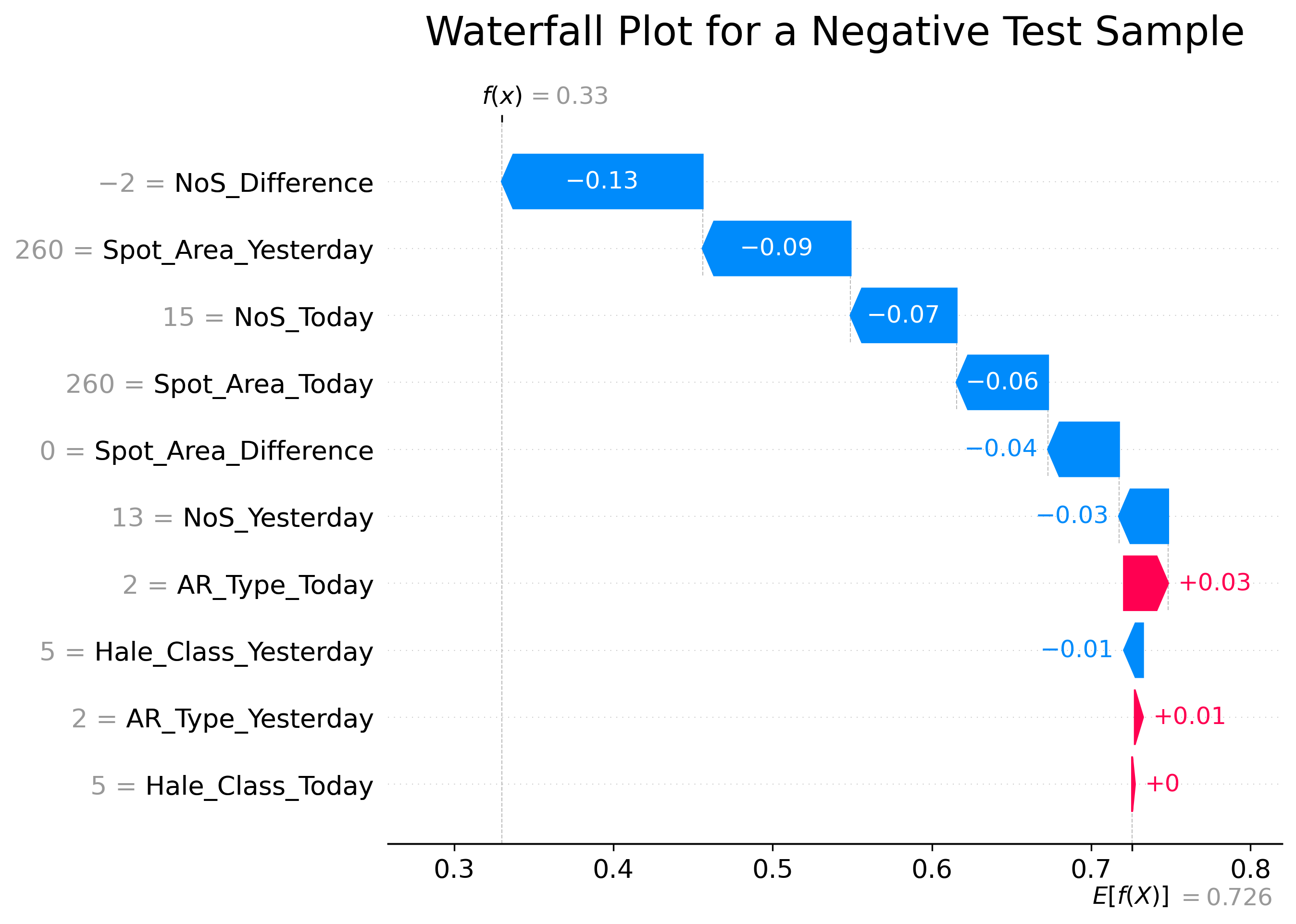}
\caption{Waterfall plot for a test sample predicted to be negative. The plot shows the relative contribution of each feature to
the model’s prediction $f(x)$ = 0.33, 
starting from the base value $E[f(x)]$ = 0.726.}
\label{fig:waterfallnegative}
\end{figure}

Figure \ref{fig:waterfallpositive} presents the waterfall plot for a test sample predicted
to be in the positive class. 
Figure \ref{fig:waterfallnegative} presents the
waterfall plot for a test sample predicted
to be in the negative class. 
These plots
display the relative contributions of the different features in
order of importance.
In Figure \ref{fig:waterfallpositive}, we see that the three most important features\\
(NOS\_Difference, 
NoS\_Yesterday,
AR\_Type\_Today)
all make positive contributions
with positive SHAP values, 
drawing the model to predict that the test sample is positive.
In contrast, in Figure \ref{fig:waterfallnegative},
the six most important features all make negative contributions with negative SHAP values, drawing the model to predict that the test sample is negative.
When comparing the waterfall plots to the three previous plots (beeswarm, bar, and decision), it
is important to recall that the three previous plots are designed for global interpretation
based on all test samples in the test set, while
the waterfall plots are designed for local interpretation inspecting specific
test samples. 
Thus, the rankings of the importance of the features between the
three previous plots
(beeswarm, bar, and decision)
and the waterfall plots are different.
However, comparing the previous three plots
(beeswarm, bar, and decision)
and
the waterfall plots can provide
valuable insight to identify possible overlaps in importance
of the features. 
NoS\_Difference, for example, is of great importance
in all four plots
(beeswarm, bar, decision, and waterfall).

\section{Discussion}
\label{sec:discussion}

\begin{figure}
\centering
\hspace*{-0.7cm}
\includegraphics[width=0.7\linewidth]{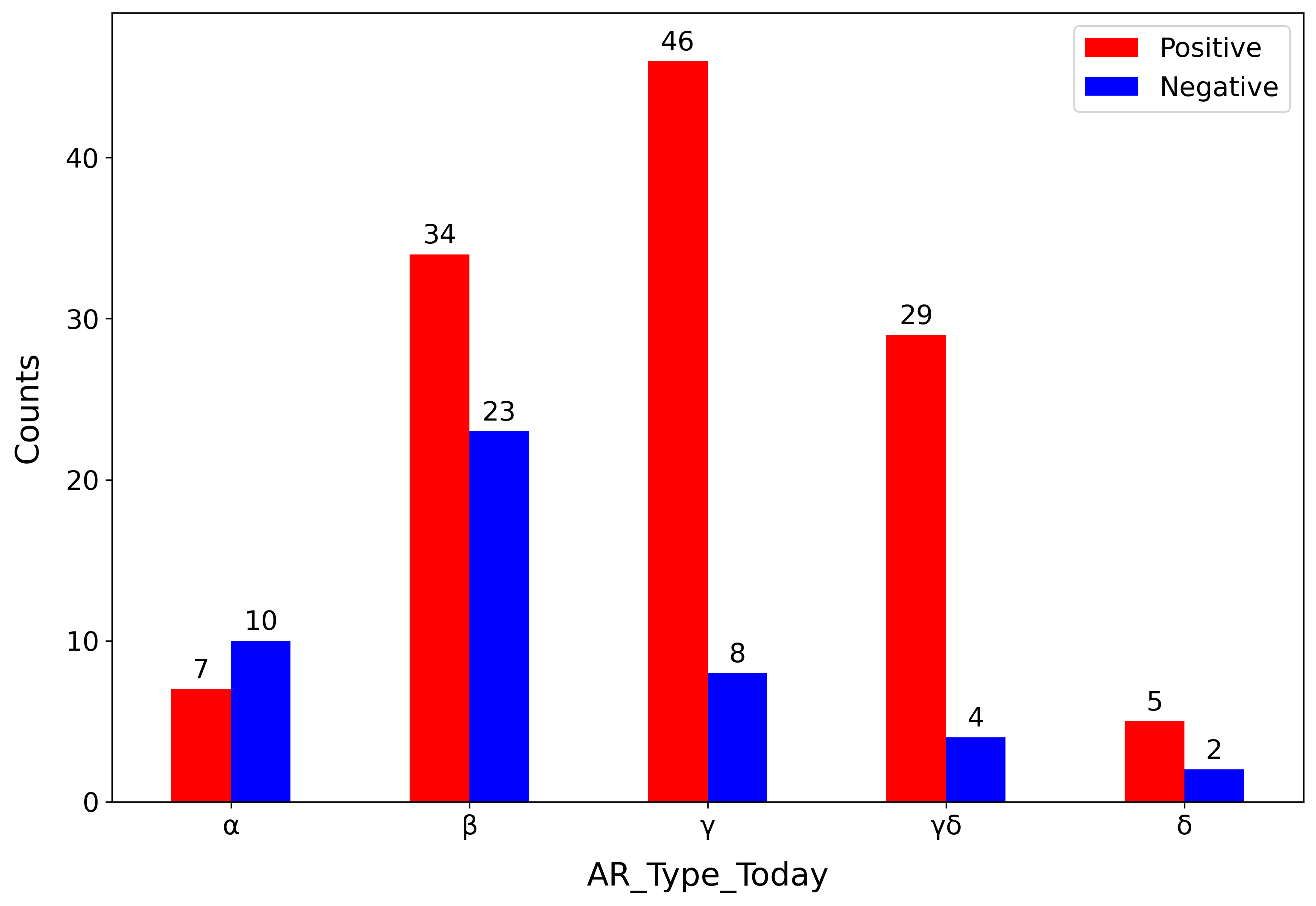}
\caption{Number of test samples for the varying feature values (categorical values) of AR\_Type\_Today obtained based on the 121 positive and 47 negative samples in the test set. For each feature value $v$, there is a clear difference between the number of positive test samples with the feature value $v$ and the number of negative test samples with the feature value $v$. This AR\_Type\_Today is the most important feature among the 10 physical features considered in this study. Our RF model would prefer to use this feature for flare classification.}
\label{fig:artypetoday}
\end{figure}

\begin{figure}
\centering
\hspace*{-0.7cm}
\includegraphics[width=0.7\linewidth]{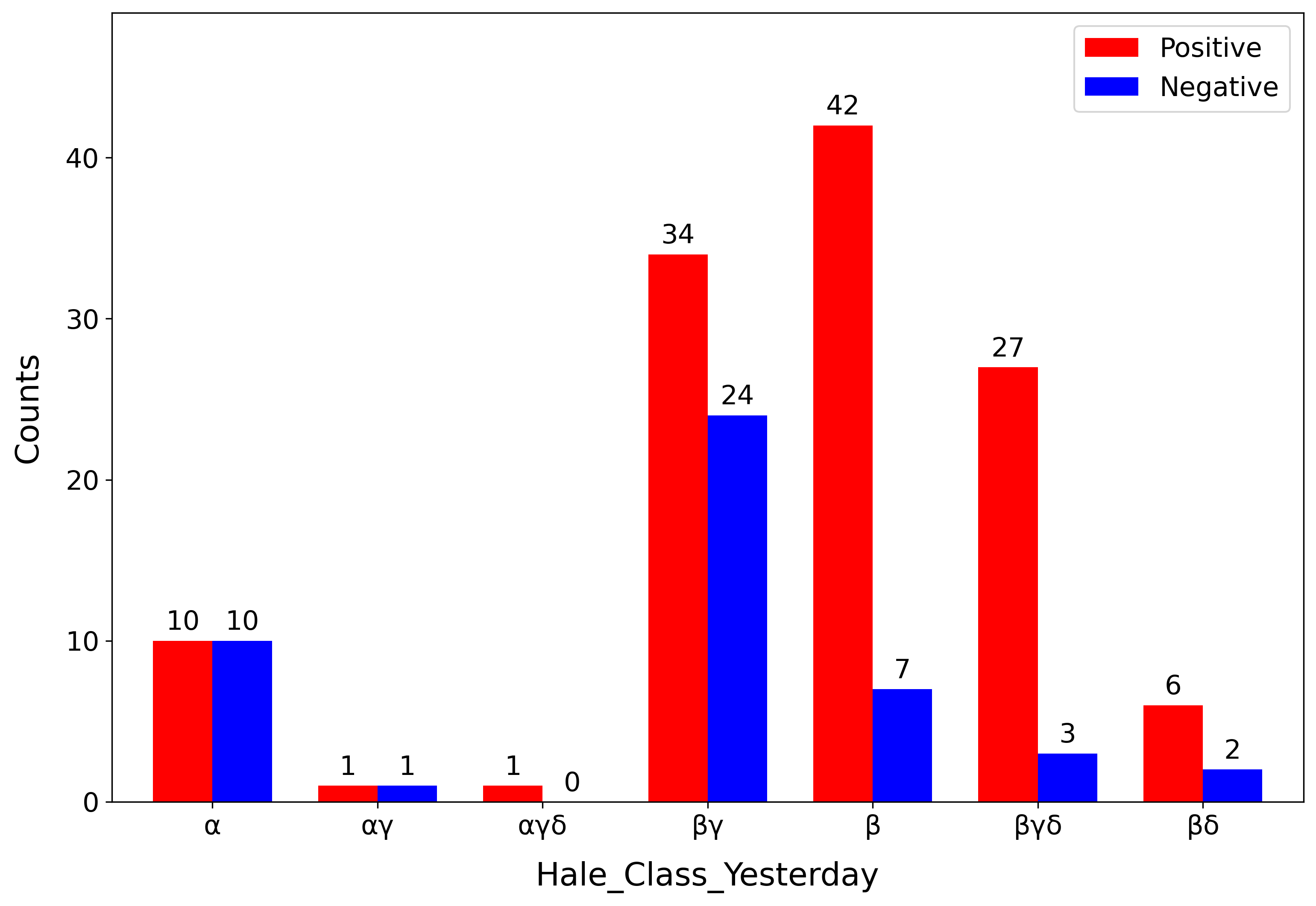}
\caption{Number of test samples for the varying feature values (categorical values) of Hale\_Class\_Yesterday obtained based on the 121 positive and 47 negative samples in the test set. No test sample has $\alpha$$\delta$, and hence this feature value is not shown in the figure. Compared to AR\_Type\_Today in Figure \ref{fig:artypetoday}, there are 3 features ($\alpha$, $\alpha$$\gamma$, $\alpha$$\gamma$$\delta$) based on which it is hard to distinguish between positive test samples and negative test samples. 
This Hale\_Class\_Yesterday is the least important feature among the 10 physical features considered in this study. Our RF model would prefer not to use this feature for flare classification.}
\label{fig:haleclassyesterday}
\end{figure}

\begin{figure}
\centering
\hspace*{-0.7cm}
\includegraphics[width=0.7\linewidth]{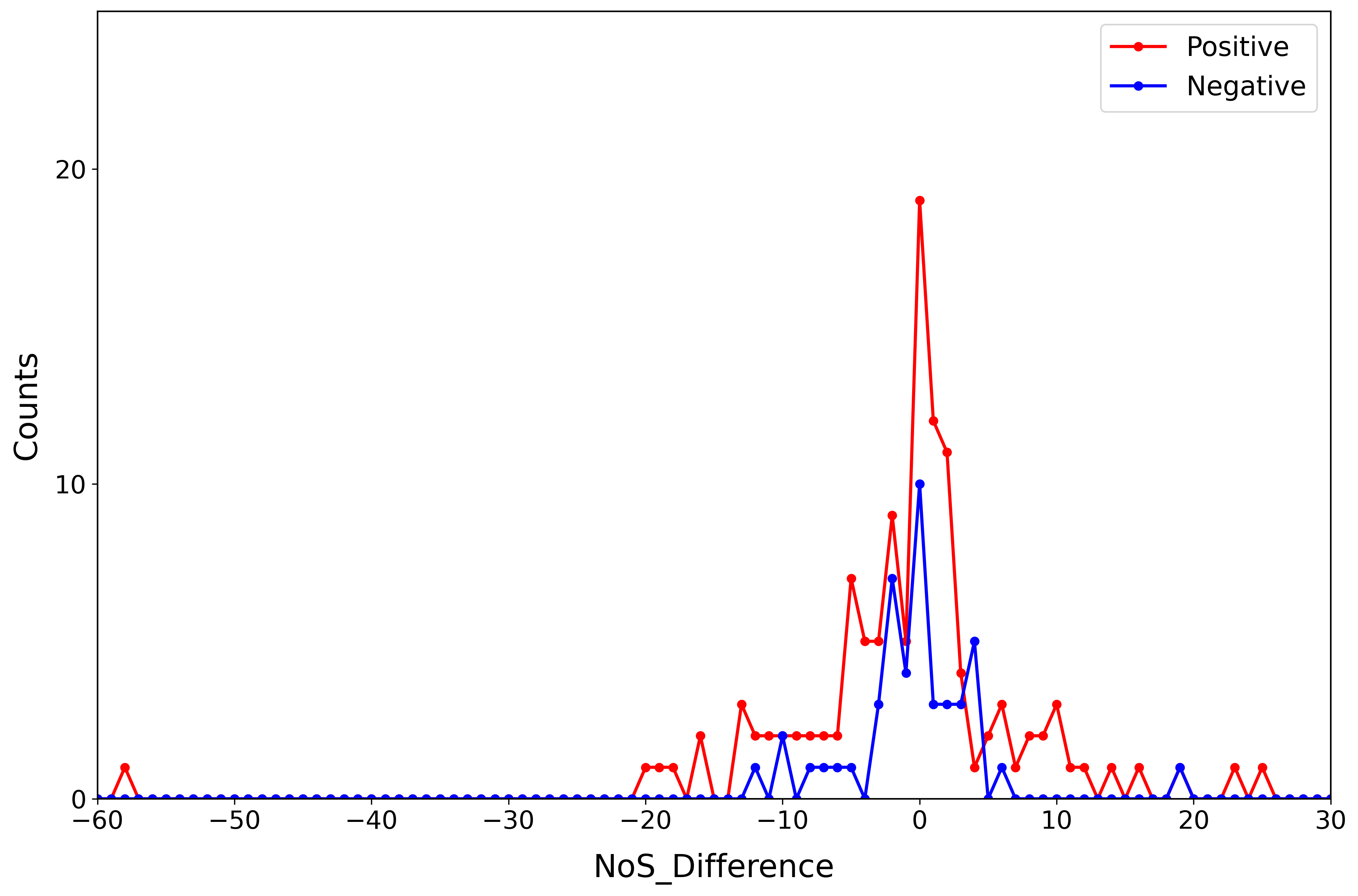}
\caption{Number of test samples for the varying feature values of NoS\_Difference obtained based on the 121 positive and 47 negative samples in the test set. The feature values are numerical values, ranging from $-58$ to 25. Like AR\_Type\_Today in Figure \ref{fig:artypetoday}, it is relatively easy to distinguish between positive test samples and negative test samples based on the feature values. This NoS\_Difference is of high importance in both global and local interpretations. Our RF model would prefer to use this feature for flare classification.}
\label{fig:nosdifference}
\end{figure}

The results in Section \ref{sec:SHAP}, show that the AR\_Type\_Today and Hale\_Class\_Yesterday features play the most important and least important roles in determining whether an AR produces a $\geq$C-class flare based on the test set. When looking at specific test samples more precisely
(i.e., the positive and negative test samples), we see that 
NoS\_Difference is of the highest importance in the decision-making process.

In the AR\_Type\_Today feature (Figure \ref{fig:artypetoday}), $\beta$, $\gamma$ and $\gamma$$\delta$ types are more than 90\% effective for positive (i.e. $\geq$C) classifications, while they have a share of more than 74\% in negative ($<$C) classifications.  If we analyze the Hale\_Class\_Yesterday feature (Figure \ref{fig:haleclassyesterday}) as the slightest factor, the $\beta$, $\beta$$\gamma$ and $\beta$$\gamma$$\delta$ classes are effective over 85\% cases in $\geq$C predictions, while they are 72\% dominant in $<$C predictions. 

More than 90\% of the 121 positive test samples (i.e. in 109 samples) have a \\
NoS\_Difference feature value between $-13$ and $10$ as seen in Figure \ref{fig:nosdifference}. The number of sunspots in the ARs was found to be unchanged for 19 cases. In more than 45\% of the samples, the number of sunspots increased, while in 39\%, the number of sunspots decreased. In the $<$C (negative) predictions, the NoS\_Difference feature takes values between $-3$ and $4$ in nearly 81\% of the data. In 10 out of 47 samples, the number of spots of ARs remained constant, while it increased in 45\% and decreased in 34\% of the negative samples. 

\section{Conclusions}
\label{sec:conclusions}

In this paper, we employ an interpretable machine learning approach to reveal the relationship between solar flares and their source active regions through a binary classification; in other words, one class is considered positive ($\geq$C class) and the other class would be negative ($<$C class). We seek to identify the physical features of a source active region that significantly influence the likelihood of
its generation of $\geq$C-class (i.e. C, M or X classes) flares. We used the Random Forest algorithm together with the SHapley Additive exPlanations (SHAP) method. The performance metric values of recall, precision, accuracy, and the F1 score were obtained as 0.81, 0.82, 0.74, and 0.82, respectively, in our model. 

SHAP decision plots for all test data provide a comprehensive view of our model's behavior, allowing us to gain deeper insights into its decision-making process
for each test sample. Today's value of AR type is obtained as the most influential, while yesterday's value of Hale class is found to be the least effective in this binary classification (Figures \ref{fig:beeswarm1} - \ref{fig:dec1}). According to a widely accepted hypothesis in the literature, as the number of sunspots increases, the flare intensity also increases. In our study, as in \cite{2004AAS...205.1002S}, the number of sunspots is not as important as stated. However, we realize that the difference between the number of spots in the AR compared to the previous day is a crucial feature in the decision-making process for positive and negative test samples, as seen in the waterfall plots (Figures \ref{fig:waterfallpositive} and \ref{fig:waterfallnegative}). When Figure \ref{fig:nosdifference} was carefully analyzed, it was found that this range of difference is between $-13$ and $10$ in 90\% of the positive class and between $-4$ and $3$ in 81\% of the negative class. 
We note that ARs of the types $\alpha$ and $\beta$ can be grouped as simple ARs, while other types were classified as ARs of medium and high complexity. \cite{10.1093/mnras/stw2742} reported that 79\% of the flares were generated by medium and high complexity AR groups in their study without binary classification. In the test sample (Figure \ref{fig:artypetoday}) of our study, it is found around 56\%. 

Our study reveals that (Figure \ref{fig:artypetoday}) 34\% and 66\%  of positive test samples are formed by simple and complex ARs, respectively. In negative test samples, these rates are 70\% and 30\% for simple structured and more complex ARs, respectively. In their study on the relationship between magnetic disturbances and flares in active regions, \citet{OLOKETUYI2023101972} reached results similar to ours. According to their results, 45\% of the flares in the positive class are formed by ARs with simple structure, while 55\% are formed by ARs with complex structure. For flares in the negative class, these rates are found to be 71\% and 29\% for simple and complex ARs, respectively. In \citet{Yang_2017}, it was reported that 46\% of $\geq$C class flares were caused by simple ARs while 54\% were caused by more complex ARs in their study of the statistical relationships between flare and AR properties. For $<$C class flares, they found that simple structured ARs are 60\% effective, while for complex structured ARs they gave a rate of 40\%. As can be seen, it would not be erroneous to conclude that the results of our study, in which a binary classification is defined as positive versus negative, are in line with the rates presented in the studies investigating the relationship between AR complexity and flares in the literature. 

Our results indicate that the $\beta$, $\gamma$ and $\gamma$$\delta$ types contribute to $\geq$C and $<$C classifications with more than 90\% and 74\% effectiveness, respectively. This result is in good agreement with the result obtained in Figure 2 of \citet{Sammis_2000}. They found that flares having X-ray fluxes from $10^{-6}$ $Wm^{-2}$ to $10^{-4}$  $Wm^{-2}$ (i.e. from C to X classes) are mostly produced by ARs having bipolar sunspot groups or more complex structures (namely ARs of $\beta$ or having higher complexity).  Although our study does not show that the surface area of ARs is effective, in a review article by \citet{2019LRSP...16....3T}, it is stated that the probability of flare eruption is proportional to the spot area and increases with the spot complexity ($\beta$, $\gamma$ and $\gamma$$\delta$ ) as found in the current study.

However, the causative mechanism for these eruptions remains unidentified. Consequently, traditional solar flare forecast fundamentally relies on the statistical correlation between solar flares and AR features derived from observational data. In this work, an interpretable machine learning approach was employed to elucidate the association between solar flares and their originating ARs using binary classification. 
The directions of future research include the following.
\begin{itemize}
\item Employing a broader spectrum of data, particularly examining data that encompass observational tools that have advanced in recent years, can reveal new findings.
\item Integrating data from contemporary technologies, such as sophisticated machine learning analysis tools, through a more diversified data set might yield innovative insights and improve predictive accuracy. 
\item Examining data over prolonged durations can uncover enduring trends and patterns that may not be evident in shorter datasets. This can aid in comprehending the evolution of the phenomenon and its possible implications.
\end{itemize}

\section*{Data Availability Statement}
The 10 physical features used in this study are taken from the Solar Monitor \\
database
(https://www.solarmonitor.org/).
The GOES solar flare catalog with flare class information
is taken from XRT Flare Catalog (https://xrt.cfa.harvard.edu/flare\_catalog/).

%\bibliography{flare}

\end{document}